\documentclass{aastex631}
\usepackage{amsmath}
\usepackage{amssymb}
\usepackage{graphicx}
\usepackage{float}

\usepackage{mathtools}  
\usepackage{xfrac}  
\usepackage{ulem}
\usepackage{hyperref}

\newcommand{\req}[1]{Eq.~({\ref{#1}})}

\newcommand{\reff}[1]{Fig.~{\ref{#1}}}

\shorttitle{Strong screening in Big Bang nucleosynthesis}

\begin{document}

\title{Self-consistent strong screening applied to thermonuclear reactions}
\shortauthors{Grayson et al.}

\author[0000-0001-9985-1822]{Christopher Grayson}
\affiliation{Department of Physics, The University of Arizona, Tucson, Arizona 85721, USA}
\email{chrisgray1044@arizona.edu}

\author[0000-0001-5038-8427]{Cheng Tao Yang}
\affiliation{Department of Physics, The University of Arizona, Tucson, Arizona 85721, USA}

\author[0000-0003-2704-6474]{Martin Formanek}
\affiliation{ELI Beamlines Facility, The Extreme Light Infrastructure ERIC, 252 41 Doln\'{i} B\v{r}e\v{z}any, Czech Republic}

\author[0000-0001-8217-1484]{Johann Rafelski}
\affiliation{Department of Physics, The University of Arizona, Tucson, Arizona 85721, USA}

\keywords{Big Bang nucleosynthesis (151), Plasma astrophysics (1261), Nuclear physics (2077), Nuclear astrophysics (1129), Nuclear fusion (2324), Nuclear reaction cross sections (2087) }

\begin{abstract}
Self-consistent strong plasma screening around light nuclei is implemented in the Big Bang nucleosynthesis (BBN) epoch to determine the short-range screening potential, $e\phi(r)/T \geq 1$, relevant for thermonuclear reactions. We numerically solve the non-linear Poisson-Boltzmann equation incorporating Fermi-Dirac statistics, adopting a generalized screening mass to find the electric potential in the cosmic BBN electron-positron plasma for finite-sized $^4$He nuclei as an example. Although the plasma follows Boltzmann statistics at large distances, Fermi-Dirac statistics is necessary when work performed by ions on electrons is comparable to their rest mass energy. While strong screening effects are generally minor due to the high BBN temperatures, they can enhance the fusion rates of high-$Z>2$ elements while leaving fusion rates of lower-$Z\le 2$ elements relatively unaffected. Our results also reveal a pronounced spatial dependence of the strong screening potential near the nuclear surface. These findings about the electron-positron plasma's role refine BBN theory predictions and offer broader applications for studying weakly coupled plasmas in diverse cosmic and laboratory settings. 
\end{abstract}

\section{Introduction}\label{sec:intro}

\subsection{The Big Bang nucleosynthesis epoch}
We address the formation of light elements during the Big Bang nucleosynthesis (BBN) epoch in the temperature range $86\,\mathrm{keV}>T>20\,\mathrm{keV}$~\citep{Pitrou:2018cgg}. In this temperature regime, electron-positron pairs ($e^-e^+$ plasma) are abundant, as highlighted in various recent studies~\citep{Wang:2010px, Hwang:2021kno, Rafelski:2023emw}. We determine the magnitude and effect of the self-consistent strong field screening for finite-sized stationary nuclei in the early Universe induced by this exotic $e^-e^+$ plasma.  These effects are of interest for precision BBN calculations at the 0.1\% level and are of general interest in exploring the properties of plasma theory.

Here, we explore various static and nonlinear models of charge screening, a collective effect within the plasma that alters the potential between nuclei.  Plasma screening involves electrons surrounding an ion's charge $Ze$ (elementary charge $e>0$), which effectively 'screens' or diminishes the influence of other nuclear charges beyond their radius, lowering the Coulomb barrier. In the context of nuclear reactions, this reduction in the Coulomb barrier facilitates increased penetration probability. This, in turn, boosts the rates of thermonuclear reactions and consequently alters the abundance of lighter elements formed in the early universe.

Traditionally, most BBN plasma studies assume the ``weak-field'' limit where the electromagnetic potential energy $\phi(r)$ is small compared to the thermal energy $T$
\begin{equation}
    \frac{q \phi(r)}{T} \ll 1\,.
\end{equation}
In this case, the movement of plasma particles is dominated by thermal fluctuations rather than the Coulomb force. Plasmas satisfying this condition are weakly coupled, indicating plasma effects will lead to linear corrections to the potential, such as in  Debye-H\"{u}ckel theory~\citep{Debye:1923}.

Due to the relatively low baryon number density $n_B$ in the early Universe~\citep{Grayson:2023flr}, the internuclear distance $a = n_B^{-1/3}$ is large and the macroscopic properties of the early Universe plasma satisfy the weak field limit for the inter-nuclear spacing $a$ because, at this distance, the potential is much weaker than the thermal energy, which is below $86\,$keV
\begin{equation}\label{eq:weakcond}
    \frac{Ze\phi(a)}{T} \ll 1\,, \quad \text{with} \quad a = n_B^{-1/3}\,,
\end{equation}
for light nuclei with charge $Z$. The weak field limit can accurately describe the electromagnetic fields in the plasma at large distances on the order of the inter-nuclear spacing $a$ but not at short distances where $Ze\phi/T$ could be larger than one.

Although the BBN plasma is weakly coupled, nonlinear corrections to the short-distance electromagnetic potential may be relevant to quantum tunneling in thermonuclear reactions. This is because the Gamow energy $E_G$ at which nuclei are most likely to tunnel is higher than the BBN thermal energy~\citep{Shaviv:1996}. The internuclear distance corresponding to Gamow energy is on the order of femtometers, such that for the short-range potential
\begin{equation}\label{eq:strongcond}
    \frac{Ze \phi(r_{E_G})}{T} > 1\,,
\end{equation}
where $r_{E_G}$ is the classical turning point at the Gamow energy. 
\begin{equation}\label{eq:gamow}
    E_G = \left(\frac{(\pi T Z_1 Z_2 \alpha)^2 \mu_r }{2}\right)^{1/3}\,.
\end{equation}
The reduced mass of the colliding light nuclei is $\mu_r$, and $Z_1$ and $Z_2$ are the respective charges of the nuclei. The condition \req{eq:strongcond} indicates that although the BBN plasma can be treated globally as weakly coupled, locally, the short-range potential can have nonlinear corrections simply because the electrostatic energy close to a nucleus can be much higher than the thermal energy. Naively using Boltzmann statistics, one expects exponential enhancement of the charge density~\citep{Grayson:2023flr}. We anticipate that these nonlinear corrections are relevant because they are on the order of the weak-field corrections due to dynamic motion and damping.

\subsection{Methods of evaluation}
Plasma screening effects were first considered in 1954 by \citep{Salpeter:1954nc}, who proposed evaluating the enhancement of nuclear reactions by employing the static Debye-H\"{u}ckel potential~\citep{Debye:1923,Salpeter:1969apj,Famiano:2016hhs}. These applications focus on collision-less plasma only. Later, this approach was generalized for nuclei moving in the plasma~\citep{Hwang:2021kno,Carraro:1988apj, Gruzinov:1997as, Opher:1999jh, Yao:2016cjs} {\it i.e.\/}, `dynamic' screening. In our previous work, we addressed scattering damping~\citep{Formanek:2021blc} in the quantum electrodynamic (QED) $e^-e^+\gamma$ plasma~\citep{Grayson:2023flr} where the BBN reaction network occurs. Similar numerical simulations were performed to study damping in the $e^-e^+\gamma$ plasma~\citep{Sasankan:2019oee,Kedia:2020xdc}.

To estimate the damped-dynamic enhancement of thermonuclear reactions during the BBN epoch, we can use the approximate analytic damped-dynamic screening potential denoted as $\phi_\text{DD}(r)$ in~\citep{Grayson:2023flr}. For  weak screening with potential $e\phi/T \ll 1$, the reaction rate enhancement is related to the usual Salpeter enhancement factor~\citep{Salpeter:1954nc}, which only depends on the value of $\phi_\text{DD}(r)$ at the origin $r=0$
\begin{equation}\label{eq:DD}
    e\phi_\text{DD}(0) = Z \alpha \left(m_D c^2 - \frac{3}{2} \beta \sigma_{0}\right)\,,
\end{equation}
where $\alpha \approx 1/137$ is the fine structure constant, $\beta$ is the thermal velocity of ions in the plasma 
\begin{equation}
    \beta = \sqrt{\frac{2 T}{M}}\,
\end{equation}
with $M$ being the mass of nuclei and $T$ is the temperature of the BBN electron-positron plasma, with $k_B = 1$. Then
$m_D$ is the Debye screening mass,  
and $\sigma_{0}$ is the static conductivity of the early universe during BBN. The Debye screening mass is related to the Debye length $\lambda_D$ as 
\begin{equation}
 m_D c^2 = \frac{\hbar c}{\lambda_D}\,,
\end{equation}
where $\hbar c = 197.3\ \mathrm{MeV}\ \mathrm{fm}$. From now on, we will absorb $c$ into $ct\rightarrow t$, such that $m_D c^2 \rightarrow m_D$ has units of energy. \req{eq:DD} is only valid in the weak damping limit $\omega_p<\kappa$, with $\kappa$ being the average rate in of collisions and $\omega_p$ the plasma frequency.
The thermal velocity of ions during BBN is very small due to their large mass $\beta < 10^{-2}$. The conductivity of the early universe $\sigma_{0}$ is estimated in~\citep{Grayson:2023flr} to be $ 0.23\,$keV. Therefore, the dynamic correction in the second term of \req{eq:DD} is small compared to the static result in the first term, $m_D \approx 2.7\,$keV. For this reason, the dynamic contribution to screening will be neglected in this work. 

Approaches to strong screening have been previously developed in astrophysical plasmas such as solar plasmas and in $^{12}$C plasmas within white dwarfs. This research began with Salpeter in 1954, modeling a strongly coupled plasma with a uniform density of electrons that can completely screen nuclei. The following papers continue this research~\citep{Salpeter:1954nc, Dewitt1973,Itoh1977,Itoh1979, Ichimaru:1982, Kravchuk:2014xsa}. Strong screening models are applicable for plasmas where the Coulomb energy is much larger than the thermal energy, even at distances larger than the ion separation. In this work, we study `intermediate screening,' which we call self-consistent strong screening since it captures both the strong and weak field regimes and involves a self-consistent determination of the potential.

The strong screening potential is described by the Poisson-Boltzmann equation~\citep{Dzitko:1995xyz, Gruzinov:1998, Bruggen:1997, Bruggen:2000, Bi:2000, Liolios:2004, Luo:2020}. The Poisson-Boltzmann equation is also used in physical chemistry to calculate the electromagnetic potential in electrolytic solutions~\citep{Fogolari:2002xyz}. Traditionally, using the Boltzmann approximation assumes point-like charges, which cause the screening potential to diverge near their locations~\citep{Gruzinov:1998}. In the past, this issue was circumvented by calculating strong screening using a cluster expansion of weak field screening~\citep{Graboske:1973xyz}, by considering ion correlations~\citep{Dewitt1973, Itoh1977}, and employing the density matrix equation in quantum statistical mechanics~\citep{Gruzinov:1998, Elsing:2022xyz}. We solve the Poisson-Boltzmann equation directly by implementing a finite-sized source charge density and using Fermi-Dirac statistics. Other solution methods to the Poisson-Boltzmann equation using Fermi-Dirac statistics are explored in the context of stellar fusion by \citep{Bruggen:2000}, where the approximate screening potential is found analytically using matching conditions. Numerical studies of the Poisson-Boltzmann equation were performed in~\citep{Cowan:1958xyz} to compare with quantum mechanical calculations. 

\subsection{Novel approaches and results}

Our study introduces several advancements and findings in plasma screening during the Big Bang nucleosynthesis (BBN) epoch. Our main result is the short-range potential around light nuclei. We introduce finite-sized nuclei into the Poisson-Boltzmann equation, eliminating singular behavior mentioned in~\citep{Gruzinov:1998} that typically arises with point-like charges. Additionally, we formulate a generalized screening mass that characterizes the strength of polarization within the plasma, providing a framework for comparing various screening models. Our results indicate that Fermi-Dirac statistics is necessary when the work performed by the external potential of ions on electrons is comparable to their rest mass energy, even though the plasma adheres to Boltzmann statistics at larger distances. We establish an analytic strong screening mass in the ultrarelativistic limit for Fermi-Dirac statistics, which could be instrumental in developing a dynamic theory of strong screening in future studies.

While strong screening is generally small due to the high temperatures prevalent during BBN, it is interesting since it enhances the fusion rates of high-$Z$ elements while leaving elements with lower-$Z$ relatively unaffected. Our findings reveal that although the overall screening effect diminishes with decreasing temperature, the relative enhancement of strong screening over weak screening grows as the temperature decreases. We also find that the strong screening potential exhibits a pronounced spatial dependence near the nuclear surface. The step nature of this potential shown in \reff{fig:pot_Comp} implies that relying solely on the screening energy near the origin can lead to an overestimate of the reaction rate enhancement due to strong screening. This overestimate was previously noted in~\citep{Itoh1977}.

In Section~\ref{NSLN}, we review the theoretical background of self-consistent strong screening. Section~\ref{sec:Solve} presents numerical Solutions to the Poisson-Boltzmann equation for $^4$He ions in the BBN plasma. In Section~\ref{sec:Enhancement}, we estimate the strong screening enhancement factor for $^4$He-$^4$He and $^{12}$C-$^{12}$C Collisions by evaluating the WKB tunneling probability of the strong screening potential. Section~\ref{sec:Conc_Disc} reviews our results and discusses their implications for BBN and fusion.

\section{Self-Consistent Strong Screening of Light Nuclei}\label{NSLN}

We find the screened potential $\phi(r)$ in plasma by solving the Poisson equation for the induced polarization charge density $\rho_\mathrm{ind}(r)$ and the external charge density of the light nuclei $\rho_\mathrm{ext}(r)$
\begin{equation}\label{eq:Poss}
    -\nabla^2\phi(r) =\rho_\mathrm{tot}(r)/\varepsilon_0= [\rho_{\mathrm{ext}}(r) +  \rho_{\mathrm{ind}}(r)]/\varepsilon_0\,,
\end{equation}
where $\varepsilon_0$ is the vacuum permittivity. The equilibrium-induced charge density is the difference between the charge density of electrons and positrons
\begin{equation}\label{eq:chargesum}
   \rho_{\mathrm{ind}}(r) =  e n_+(r) - e n_-(r) \,,
\end{equation}
where $n_\pm(r)$ represents the number density of electrons and positrons.
The induced charge density can be calculated by assuming some statistical distribution for charges in the plasma. For the choice of Boltzmann statistics, one recovers the usual Poisson-Boltzmann equation.

We introduce the re-scaled potential
\begin{equation}\label{eq:indchgB}
    \Phi(r)  \equiv \frac{e\phi(r)}{T}\,,
\end{equation}
to create a dimensionless variable $\Phi$ which compares the Coulomb energy $e\phi$ to the plasma temperature $T$. The re-scaled Poisson equation reads
\begin{equation}\label{eq:Poss2}
    -\nabla^2\Phi(r) - e\rho_{\mathrm{ind}}(r)/(\varepsilon_0 T)= e\rho_{\mathrm{ext}}(r)/(\varepsilon_0 T)\,.
\end{equation}
We introduce the re-scaled external charge distribution $P_{\mathrm{ext}}$ to simplify the right-hand side 
\begin{equation}\label{eq:Pext}
 P_{\mathrm{ext}}(r)  \equiv e \frac{\rho_{\mathrm{ext}}(r)}{\varepsilon_0 T} = \frac{4 \pi Z \alpha \hbar c}{\pi^{3/2}R^3T}e^{-\frac{r^2}{R^2}}\,,
\end{equation}
where we chose to model the charge distribution of a nucleus as a Gaussian and $R$ is the root-mean-squared charge radius
\begin{equation}
    R = \sqrt{\frac{2}{3}} R_\alpha\,.
\end{equation}
For the radius of $^4$He we use the charge radius $R_\alpha = 1.67824\,$fm~\citep{Krauth:2021foz}. For the radius of $^{12}$C, we used an equilateral cluster of alphas \citep{smith2020hoyle}.
We rewrite the Poisson equation \req{eq:Poss} in terms of an effective screening mass to make the screening strength more explicit. The screening mass is defined as
\begin{equation}\label{eq:mscreen}
\frac{m_s^2(\Phi)}{(\hbar c)^2} \equiv -\frac{ e\rho_{\mathrm{ind}}(r)/(\varepsilon_0 T)}{ \Phi(r)}\,. 
\end{equation}
Using this screening mass \req{eq:Poss2} takes the form
\begin{equation}\label{eq:PossBoltz}
    -\nabla^2\Phi(r) + \frac{m_s^2(\Phi)}{(\hbar c)^2}\Phi(r) =  P_{\mathrm{ext}}(r)\,.
\end{equation}
The screening mass $m_s^2$ is related to the usual Deybe mass but exhibits additional nonlinear behavior due to its dependence on the potential. Using this generalized screening mass, we will compare Boltzmann and Fermi-Dirac equilibrium distributions for the plasma.

\subsection{Boltzmann Statistics}\label{sec:Boltz}
The full equilibrium Boltzmann distribution necessary to describe the short-range potential is~\citep{Hakim:1967prd, DeGroot:1980dk}
\begin{equation}\label{eq:Boltz}
     f_B^\pm(x,p) = e^{-\left\{u_{\mu}[p^{\mu}+q A^{\mu}(x)]\right\}/T}\,.
\end{equation}
The charge of a plasma particle is $q$, $A^\mu(x)$ is the electromagnetic 4-potential, and $T$ is the plasma temperature. During the BBN epoch $86\,\mathrm{keV}>T>20\,\mathrm{keV}$, the number of electrons and positrons are almost equal to each other due to the charge neutrality~\citep{Grayson:2023flr}. In this temperature range, we set the chemical potential to zero for our study. We assume the plasma is at rest such that its 4-velocity reads $u^\mu = (1,0,0,0)$, then
\begin{equation}
    u_{\mu}A^{\mu}(x) = \phi(x)\, .
\end{equation}
The potential term in \req{eq:Boltz} that accounts for the energy change due to the rest frame potential (or equivalently uses the kinetic vs canonical momentum) to describe the rest energy of electrons and positrons in the plasma~\citep{Hakim:1967prd, DeGroot:1980dk}. This phase space density leads to a number density that is the normal expression but enhanced by the exponential factor in potential
\begin{equation}\label{eq:denEnhan}
    n_\pm(x) = 2\int \frac{d^3\boldsymbol{p}}{(\hbar c)^3 (2 \pi)^3 p^0} p^0 f_B^{\pm} (x,p)= n_{\mathrm{eq}} e^{- q u\cdot A(x)/T}\,,
\end{equation}
where the equilibrium density is defined as
\begin{equation}
    n_{eq} \equiv 2\int \frac{d^3\boldsymbol{p}}{(\hbar c)^3 (2 \pi)^3 }  e^{-u\cdot p/T}\,.
\end{equation}
The induced equilibrium charge density is used to calculate the screening mass \req{eq:mscreen}.

Calculating the induced charge density for the Boltzmann case using \req{eq:chargesum} and \req{eq:denEnhan} and then inserting them into the definition for the screening mass \req{eq:mscreen}, one finds 
\begin{equation}\label{eq:MassBoltz}
   m_{s,\mathrm{Boltz}}^2(\Phi)= m_D^2 \frac{\sinh\left[\Phi(r) \right]}{\Phi(r)}\,,
\end{equation}
since the exponential factors in \req{eq:denEnhan} combine to give hyperbolic sine. In this expression, the Debye screening mas $m_D$ has its standard definition
\begin{equation}
  \frac{1}{\lambda_D^2} = \frac{m_D^2}{(\hbar c)^2} = \frac{8 \pi \alpha \hbar c}{T} \, n_{\mathrm{eq}}\,.
\end{equation}
$\lambda_D$ is the characteristic distance over which the linearized field is screened.

\subsection{Fermi-Dirac Statistics} \label{sec:Fermi}
We anticipate that the induced charge density near the nuclear surface where $e\phi \geq m_e$ may become large enough for degeneracy effects to become important. To model this correctly, we use the relativistic Fermi-Dirac distribution
\begin{equation}\label{eq:Fermi}
    f_F^\pm(x,p) = \frac{1}{e^{\left[u_{\mu}(p^{\mu}+q A^{\mu}(x))\right]/T}+1}\,.
\end{equation}
We then sum the induced charge densities of electrons and positrons as in \req{eq:chargesum}. After simplification, one finds
\begin{equation}\label{eq:indchgF}
    m^2_{s,\mathrm{FD}}(\Phi) = \frac{8 \pi \alpha }{ T \Phi (r)}   \int  \frac{d^3\boldsymbol{p}}{(2 \pi)^3 } \frac{\sinh\left[\Phi(r) \right]}{\cosh\left(p_0/T\right)+\cosh\left[\Phi(r) \right]}\,,
\end{equation}
which does not have a simple analytic form in the full relativistic limit due to the dependence on the re-scaled potential $\Phi$~\citep{Frolov:2023zkf}. An analytic solution to \req{eq:indchgF} exists in the ultrarelativistic limit where one can approximate $p_0 = \sqrt{m^2+\boldsymbol{p}^2} \approx |\boldsymbol{p}| $ or equivalently $m/T \ll 1$. \req{eq:indchgF} is then integrated analytically into the known form~\citep{Elze:1980er} valid to all orders in $\Phi(r)$
\begin{equation}\label{eq:ultrarel}
\begin{split}
    m^2_{s,\mathrm{FD}}(\Phi) &\approx \frac{4\alpha T^2}{3 \pi} \left[\pi^2 + \left(\Phi(r)+\frac{\mu}{T} \right)^2\right]\,.
    \end{split}
\end{equation}
This is the usual ultra-relativistic Debye mass plus a quadratic dependence in $\phi$. This screening mass has much slower quadratic growth than the exponential dependence in $\Phi$ predicted by the Boltzmann distribution \req{eq:MassBoltz}.

Keeping higher order terms in a low density expansion $z = {m}/({ \Phi T})\ll 1$ [see \citep{Birrell:2024bdb}] and for $\mu=0$  one finds~\citep{Kodama:2002}
\begin{equation}
     m^2_{s,\mathrm{FD}}(\Phi) \approx \frac{4\alpha}{3 \pi} \left( \frac{\pi^2}{2} T^2\frac{2-z^2}{\sqrt{1-z^2}} + (\Phi T)^2(1-z^2)^{3/2} +\frac{7 \pi^4 T^4}{40 (\Phi T)^2}\frac{z^4}{(1-z^2)^{5/2}}\right)\,.
\end{equation}

\begin{figure}[h]
    \includegraphics[width=0.95\linewidth]{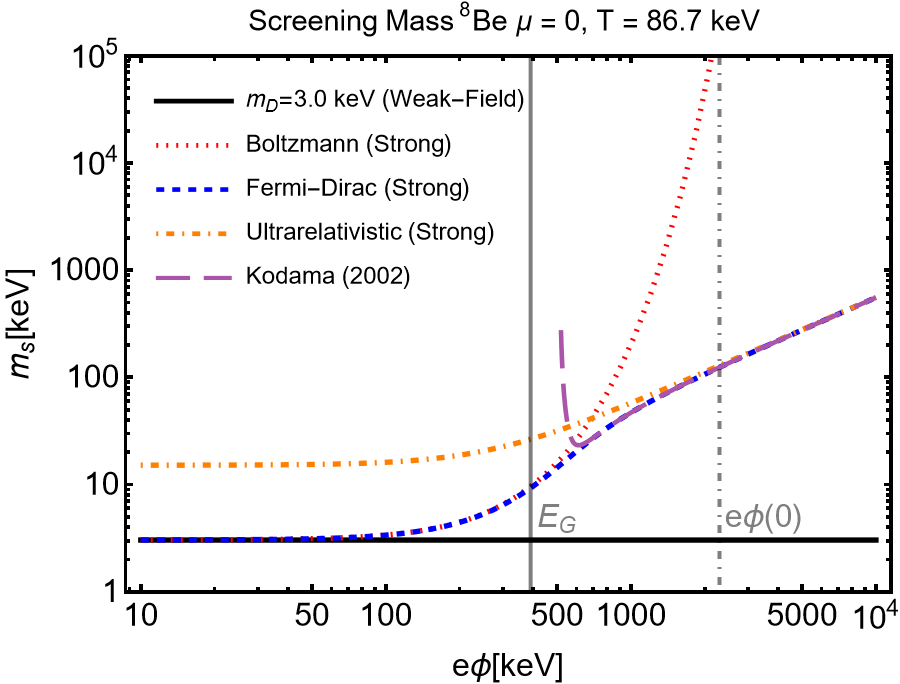}
    \caption{ The effective screening mass for both Fermi \req{eq:Fermi} and Boltzmann \req{eq:Boltz} statistics are shown as a blue-dashed line and a red-dotted line, respectively. We also show the ultrarelativistic expansion as an orange dashed-dotted line and the series expansion derived in~\citep{Kodama:2002} shown as a purple long-dashed line. Strong screening predicts a much larger screening effect at large $\Phi = e\phi/T$ values than weak screening. The Gamow energy $E_G \approx 390$ keV, the most probable tunneling energy, is shown as a gray vertical line, and the value of the potential at the origin $e\phi(0) \approx 2300\,$keV is shown as a dashed gray line. }
    \label{fig:Denisty_Comp}
\end{figure}

In \reff{fig:Denisty_Comp}, we compare the effective screening mass \req{eq:mscreen} for a Fermi-Dirac distribution and a Boltzmann distribution for varying values of the potential $\phi$. In general, a larger screening mass indicates more screening charge will be present for a particular external potential $\phi$, leading to a larger enhancement of nuclear reaction rates. In \reff{fig:Denisty_Comp}, the Boltzmann distribution vastly over-predicts the screening mass and, thus, the screening effect. This is because it does not include the stacking of fermion states of electrons and positrons in the polarization cloud. The domain of the screening mass relevant for quantum tunneling lies between the gamow energy and the potential value near the origin $E_G< e\phi< e\phi(0)$. In \reff{fig:Denisty_Comp}, we see that in this region, strong screening predicts around a $10>m^2_{s,\mathrm{FD}}/m_D>100$ times larger screening mass as compared to weak screening. This is a much larger change in the screening effect than predicted by dynamic collision-less screening~\citep{Hwang:2021kno} and damped dynamic screening~\citep{Grayson:2023flr}. For large values of potential $\phi$, the Fermi-Dirac screening mass $m_{s,\mathrm{FD}}$ approaches the ultrarelativistic limit \req{eq:ultrarel}. This indicates that the analytic expression for the ultra-relativistic Fermi-Dirac screening mass fully captures the short-distance potential and the behavior of a high-temperature strong field plasma in general. This could potentially lead to huge simplification of analytic methods. To numerically solve \req{eq:PossBoltz} we must use the relativistic Fermi-Dirac screening mass \req{eq:indchgF} instead of the analytic result \req{eq:ultrarel}, since our boundary conditions are given at large distances where \req{eq:ultrarel} is invalid.

\section{Solving the strong field Poisson equation self consistently}\label{sec:Solve}

To find the numerical solution to \req{eq:PossBoltz} we start the integration of \req{eq:PossBoltz} at large values of $r$ where the weak field condition \req{eq:weakcond} applies since there we know the analytic solution [see Appendix \ref{sec:freechg}, \req{eq:Stat_Vac}] up to an overall normalization due to the additional screening that will occur at short ranges. For this calculation, we typically began integration at 1 \AA, using the weak field analytic solution,
\begin{equation}\label{eq:Qeff}
\Phi_{r \gg \hbar c/m_s} (r) \approx \frac{Q_{\mathrm{eff}} \,e^{-m_D r}}{r}\,.
\end{equation}
as a boundary condition at large distances [for better accuracy we use the Debye-H\"uckel field for Gaussian chargers \req{eq:Stat_Gauss}].
One can integrate inwards with any standard solver to small values of $r$ varying a shooting parameter $Q_{\mathrm{eff}}$, which represents the effective charge at large distances. The solution will either diverge to positive or negative infinity based on whether the effective charge has been chosen to be too large or too small. One can then iterate the shooting algorithm, increasing or decreasing the effective charge seen at large distances until the divergence point converges sufficiently close to $r=0$.

The numerical solution can be checked for consistency as discussed in Appendix \ref{sec:check}. However, the precision of the solution was mainly determined by ensuring the stability of the solution when varying parameters such as step size, boundary conditions, and initial guesses for the shooting algorithm. The overall accuracy of the solution is determined by the choice of boundary conditions, which we assume is the analytic solution at large distances \req{eq:Stat_Gauss}.

For the Fermi-Dirac case, there is additional complexity since the density cannot be easily integrated analytically over momentum $p$. However, since the dependence on the potential is trivial, one can parameterize a numerical solution for various values of potential $\Phi$ and then use the resulting parameterization in the Poisson equation \req{eq:PossBoltz}. This parameterization is plotted in \reff{fig:Denisty_Comp}.
\begin{figure}[h]
    \includegraphics[width=0.95\linewidth]{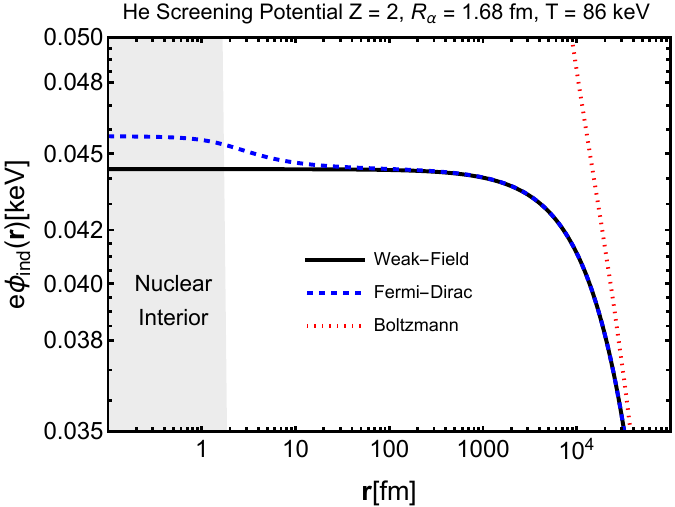}
    \caption{ The potential $e\phi_{\text{ind}}$ due to the induced screening charge density for both Fermi \req{eq:Fermi} and Boltzmann \req{eq:Boltz} statistics, including strong screening shown as a blue dashed line and a red dotted line respectively. The black solid line is the screening potential for the weak-field limit \req{eq:Stat_Gauss}. This calculation was done at temperature $T=86\,$keV at the beginning of BBN when the screening effect is largest. The gray area shows the nuclear interior at a radius of $R = \sqrt{2/3}R_\alpha$. The potential calculated using Boltzmann statistics levels off near this radius at $e\phi_\text{ind}=170\,$keV, outside the scope of this plot.}
    \label{fig:pot_Comp}
\end{figure}
\begin{figure}[h]
    \includegraphics[width=1.0\linewidth]{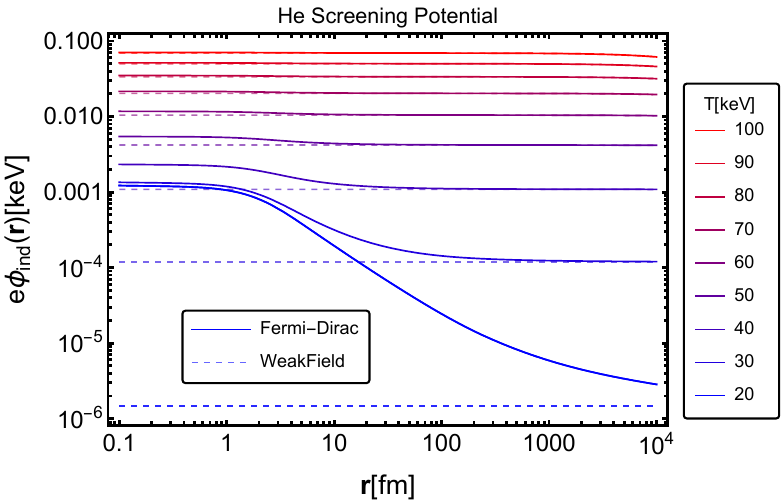}
    \caption{ The potential $e\phi_{\text{ind}}$ due to the induced screening charge density for Fermi-Dirac strong screening \req{eq:Fermi} at various temperatures as solid lines ranging from blue to red. The weak screening potentials are shown as dashed lines ranging from blue to red. Overall screening decreases with temperature $T$, but the difference between weak and strong becomes larger for small $T$.}
    \label{fig:alpha_temp}
\end{figure}

We are interested in the potential due to screening charges $\phi_{\text{ind}}$ where
\begin{equation}
\phi_\text{ind}(r)\equiv \phi(r)-\phi_{\text{vac}}(r)\,,
\end{equation} 
with $\phi$ being the total potential and $\phi_{\text{vac}}$ the potential in vacuum. This quantity is plotted for an alpha particle ($^4$He) in \reff{fig:pot_Comp}. Strong screening predicts a larger induced screening potential due to the enhancement of screening charge density \req{eq:denEnhan}. This effect is more pronounced near the origin where $e\phi/T$ is large. For an alpha particle and $T=86\,$keV, this value is around $e\phi(0)/T \approx 27$. The Boltzmann screening potential, shown as a red dotted line, calculated using \req{eq:MassBoltz} over-predicts the screening effects by a factor of up to $10^3$ since the density quickly becomes large enough that a Fermi-Dirac distribution is required to describe it correctly. The Fermi-Dirac screening potential, shown as a blue dashed line, displays the usual step behavior. The black solid line shows the weak screening potential \req{eq:Stat_Gauss}, which is nearly constant up to large distances and equal to its value at the origin
\begin{equation}
    e\phi_\text{weak}(0) = Z \alpha m_D\,.
\end{equation}
The overall screening effect in the potential \reff{fig:pot_Comp} is much less than what is naively expected by \req{eq:mscreen} since the potential at the origin is related linearly to the screening mass in the weak screening limit. In strong screening, the screening mass in \req{eq:mscreen} and plotted in \reff{fig:Denisty_Comp} can be used to accurately predict the induced charge density $\rho_\text{ind}$ but not the potential, which can only be found after solving \req{eq:PossBoltz}.

In \reff{fig:alpha_temp}, the screening potential $e\phi_{\text{ind}}$ is plotted at decreasing temperatures, the BBN temperature range being $T = 86.2 - 50\, $keV~\citep{Pitrou:2018cgg}. Temperatures lower than $20\,$keV are not considered because, at this temperature in the early Universe, positrons begin to disappear~\citep{Grayson:2023flr}. While the overall strength of screening decreases with temperature, the difference between weak and strong screening is much more pronounced at low temperatures, shown in blue, due to the large value of $e\phi/T$ near the origin. Thus, it is much more important to consider the effect of strong screening at low temperatures.

\section{Enhancement factor for nuclear reaction rate}\label{sec:Enhancement}

The thermonuclear reaction rate for a reaction process involving two nuclei is given by~\citep{Rose1998}
\begin{equation}
    R_{12} = \chi n_{eq}^1  n_{eq}^2 I
\end{equation}
where $\chi$ is a symmetry factor corresponding to $1/2$ for identical particles and $1$ for distinguishable, $ n_{eq}^1$ and $ n_{eq}^2$ are the equilibrium densities of the reacting light nuclei, which in the early universe we can assume to follow Boltzmann distribution since their mass is much larger than the temperature which is in the range of $86-50\,$keV~\citep{Pitrou:2018cgg}. The probability of the reaction occurring is determined by the integral
\begin{equation}\label{eq:integral}
    I = \sqrt{\frac{8}{\mu_r \pi T^3}}\int_{E_{\mathrm{th}}}^\infty dE \, \sigma(E) E \exp{\left( -E/T\right)}\,,
\end{equation}
where $\sigma(E)$ is the reaction cross-section, $\mu_r$ is the reduced mass of the two colliding ions, $E_{th}$ is the threshold energy for the reaction, and we assume a Boltzmann distribution for the relative energy $E$ of the collision. In a plasma environment, the screened cross-section is 
\begin{equation}\label{eq:Sfactorscreen}
 \begin{split}
\sigma_\text{sc}(E) &=\!\frac{S(E)}{E} \exp{\left( \! - \frac{2\sqrt{2 \mu_r}}{\hbar c}\!\!\int_{R}^{r_c}  \!\!\!  dr \sqrt{U_\text{sc}(r)\!- \! E}\right)}
\,.
\end{split}
\end{equation}
where $S(E)$ is the usual tabulated S-factor, which removes the $1/E$ dependence of the cross-section and the Coulomb penetration probability from the cross-section. The tunneling probability through the Coulomb barrier for the s-wave potential from the classical turning point $r_c$ to the nuclear radius $R$ is given by the WKB approximation. $U_\text{sc}(r)$ is the inter-nuclear potential energy of the screened Coulomb potential.

We are interested in finding the effect of the screened reaction rate involving the integral
\begin{equation}\label{eq:ScreenInt}
    I_\text{sc} = \sqrt{\frac{8}{\mu_r \pi T^3}}\int_{E_{\mathrm{th}}}^\infty dE S(E) e^{-g_\text{sc}(E)}\,,
\end{equation}
where
\begin{equation}\label{eq:gsc}
    g_\text{sc}(E) \equiv  E/T + \frac{2\sqrt{2 \mu_r}}{\hbar c}\int_{R}^{r_c}dr \sqrt{U_\text{sc}(r)-E}\,.
\end{equation}
To calculate the integral in \req{eq:ScreenInt}, we use the usual saddle point approximation to isolate the contribution of Coulomb penetration probability.
\begin{equation}\label{eq:ScreenIntdef}
    I_\text{sc} \approx \sqrt{\frac{16}{\mu_r T^3 g_\text{sc}''(E_G)}} S(E_G) e^{-g_\text{sc}(E_G)}\,,
\end{equation}
with Gamow energy $E_G$ defined in \req{eq:gamow}. We numerically calculate \req{eq:ScreenInt} using the saddle point approximation and compare it to various screening potential energy models, specifically weak screening to strong screening. We do not compare to Boltzmann statistics since it vastly over-predicts the screening effect.

The potential energy between the two nuclei is related to the potential $\phi$ calculated in Sect. \ref{sec:Solve} by the self energy
\begin{equation}
    U_\text{sc}(r) = \frac{1}{2}\int d^3r' \rho(r') \phi(r-r') - U(r\rightarrow \infty) = \frac{1}{2}\int d^3r'\left[\rho_1(r')\phi_2(r-r')+
    \rho_2(r')\phi_1(r-r')\right] \,.
\end{equation}
Where we have subtracted the in-plasma self-energies of the two nuclei separated at $r\rightarrow\infty$
\begin{equation}
    U(r\rightarrow \infty) = \frac{1}{2}\int d^3r'\left[\rho_1(r')\phi_1(r-r') +\rho_2(r')\phi_2(r-r')\right] \,.
\end{equation}
The total charge density is the sum of the induced plus the external charge
\begin{equation}
    U_\text{sc}(r) = \frac{1}{2}\int d^3r'\left\{\left[{\rho_\text{ext}}_1(r')+{\rho_\text{ind}}_1(r') \right]\phi_2(r-r')+ \left(1 \leftrightarrow 2\right) \right\} \,.
\end{equation}
Then, we use \req{eq:mscreen} to rewrite the induced charge density in terms of the total potential
\begin{equation}\label{eq:UscFull}
    U_\text{sc}(r>R) = \frac{1}{2} Z_1 \phi_2(r)  + \frac{1}{2}\int d^3r'\left[ m_s^2(\phi_1)\phi_1(r')  \phi_2(r-r') \right] + \left(1 \leftrightarrow 2\right)  \,.
\end{equation}
We approximate $\rho_\text{ext}$ as a point charge at distances larger than $R$, relevant for the tunneling integral in \req{eq:Sfactorscreen}.
Often, in the weak screening limit, one only considers the charge of colliding nuclei with the total potential of the other. If the charges are approximately symmetric, the screening potential is then
\begin{equation}\label{eq:potapprox}
     U_{\text{sc}}(r) \approx Z_2 \phi_1(r)\,.
\end{equation}
Since we only want to compare weak screening to strong screening, for simplicity, we neglect the second term in \req{eq:UscFull}. This approximation is only good for small $Z$ when the value of $m_s$ is much smaller than the vacuum potential. To calculate the exact potential energy in this model one would need to calculate the screening of two nuclei as they approach since the the principle of superposition does not apply when nonlinear screening is present. In the weak screening limit, the second term in \req{eq:UscFull} is known to lead to a factor $3/2$ in the Salpeter's usual enhancement factor \req{eq:weakenhance}~\citep{Bruggen:1997}. 

We will compare \req{eq:ScreenInt} to the same integral in vacuum, assuming the electric potential used to find the potential energy is given by \req{eq:Stat_Vac}
\begin{equation}
    I_\text{vac} = \sqrt{\frac{8}{\mu_r \pi T^3}}\int_{E_{\mathrm{th}}}^\infty dE S(E) e^{-g_\text{vac}(E)}\,,
\end{equation}
where
\begin{equation}
    g_\text{vac}(E) \equiv  E/T + \frac{2\sqrt{2 \mu_r}}{\hbar c}\int_{R}^{r_c}dr \sqrt{U_\text{vac}(r)-E}\,. 
\end{equation}
In this expression, $U_\text{vac}$ is the vacuum internuclear potential. We represent the enhancement of the thermonuclear reaction rate as 
\begin{equation}\label{eq:ratio}
    \mathcal{F}_\text{sc} \equiv \frac{R_\text{sc}}{R_\text{vac}}=\frac{I_\text{sc}}{I_\text{vac}}\,.
\end{equation}
In Appendix \ref{sec:finte}, we mention the effect of finite size on reaction rates, which is much larger than screening and essential to include.

When the Debye length $\lambda_D$ is very large compared to classical turning point $r_c$, the screening charge density can be approximated to be constant near the origin. In that case, the screening effect can be described as a shift in the energy of the reaction rate leading to a simple exponential enhancement referred to as the Salpeter enhancement factor given by the change in the potential energy at the origin.
\begin{equation}\label{eq:weakenhance1}
   \mathcal{F}_{\text{sc}}(0) = \exp{\left( \frac{U_{vac}(0)-U_\text{sc}(0)}
   {T}\right)} 
    = \exp{\left( \frac{H(0)}
   {T}\right)} \, .
\end{equation}
$H(0)$ is often used in literature to represent the electric potential energy of the induced charge density. Alternative derivations of \req{eq:weakenhance} lead to different forms of the Slapeter enhancement factor, summarized in~\citep{Bahcall:2002xyz}. Since we compare the enhancement factor for strong screening to weak screening, the exact form of $\mathcal{F}_{\text{sc}}$ is not important since multiplicative factors will cancel. For the weak screening limit using only the leading order potential energy \req{eq:potapprox}, one finds~\citep{Salpeter:1954nc}
\begin{equation}\label{eq:weakenhance}
\begin{split}
   \mathcal{F}_{\text{weak}}(0)  = \exp\left(\frac{Z_1 Z_2 \alpha m_D}{T}\right)\, .
\end{split}
\end{equation}


\begin{figure}[h]
    \includegraphics[width=0.95\linewidth]{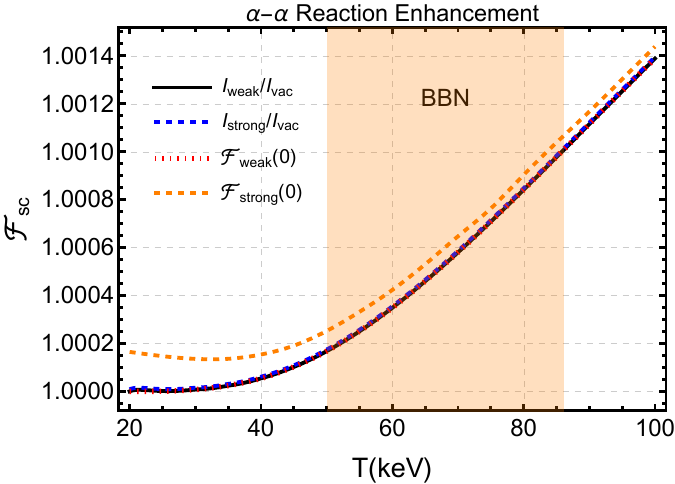}
    \caption{Reaction rate enhancement $\mathcal{F}_\text{sc}$ [see \req{eq:ratio}] for $^4$He-$^4$He scattering  as a function of temperature. The weak screening model is shown as a black solid line, and the strong screening model is plotted as a blue dashed line. The approximate enhancement factors using the screening potential energy at the origin \req{eq:weakenhance} are shown as a red dotted line for weak screening and an orange dashed line for strong screening. The orange shaded region marks the BBN temperature range $T = 50 - 86.2\, $keV.} 
    \label{fig:enhanceAlpha}
\end{figure}



In \reff{fig:enhanceAlpha}, we calculate the enhancement due to screening in the BBN plasma environment by numerically calculating the ratio $\mathcal{F}_\text{sc}$ from \req{eq:ratio} using the weak the screening potential \req{eq:Stat_Gauss} shown as a black solid line and the strong screening potential, found by solving \req{eq:PossBoltz} with Fermi-Dirac statistics, shown as a blue dashed line. This is done by calculating the WKB tunneling integral in \req{eq:gsc} using the numerical solutions for the potential plotted in \reff{fig:alpha_temp} and then estimating the integral \req{eq:ScreenInt} using the saddle point approximation \req{eq:ScreenIntdef}. The blue dashed line and the black solid line include the finite size of the colliding nuclei and the screening potential's spatial dependence. In the calculation of \req{eq:ScreenIntdef}, the change in the Gamow energy due to screening is calculated for completeness, but, in general, this change is negligible since screening does not change the distribution of ions or the penetration probability significantly.

Strong screening generally predicts a larger enhancement of reaction rates. However, for $^4$He-$^4$He collisions, this enhancement is very small, as indicated by the blue dashed line almost tracing the black solid line. This is expected since predicted  $^4$He abundances match theoretical measurements~\citep{Pitrou:2018cgg}. The discrepancy between weak and strong screening becomes a constant at high temperatures $\sim 10^{-6}$, indicating the polarization density is approaching the ultrarelativistic limit of Fermi-Dirac statistics \req{eq:ultrarel}.

The Salpeter enhancement factor $\mathcal{F}_{\text{weak}}(0)$ reasonably approximates the enhancement due to weak screening, as seen by the red dotted line tracing the black solid line. The Salpeter enhancement factor for strong screening is shown as an orange dashed line in \reff{fig:enhanceAlpha} using the value of the strong screening potential in \reff{fig:alpha_temp} at the origin. One can see in \reff{fig:enhanceAlpha} that the Salpeter approximation for screening $\mathcal{F}_{\text{strong}}(0)$ overestimates the numerical solution. This is because the induced charge density is not constant near the origin but has a step in the potential created by the Fermi function in the calculation of the strong screening potential. This overestimation of the Salpeter enhancement factor is also found in~\citep{Itoh1977}. The Salpeter enhancement factor is a less accurate approximation at low temperatures where the jump in potential is much more pronounced.


\begin{figure}[h]
    \includegraphics[width=0.95\linewidth]{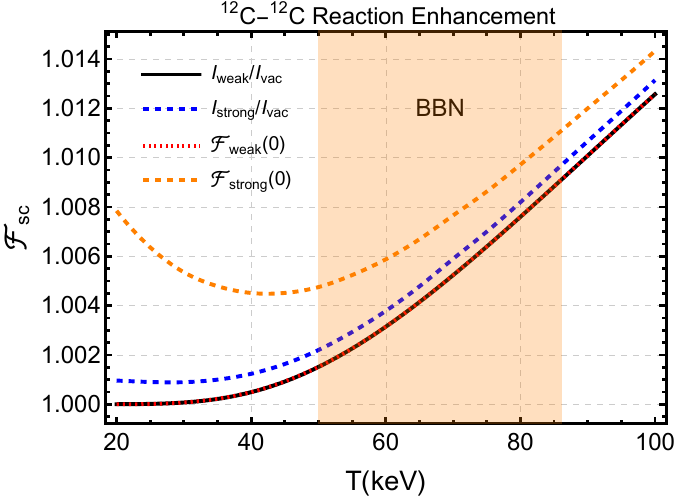}
    \caption{Reaction rate enhancement $\mathcal{F}_\text{sc}$ [see \req{eq:ratio}] for $^{12}$C-$^{12}$C scattering as a function of temperature. The weak screening model is shown as a black solid line, and the strong screening model is plotted as a blue dashed line. The approximate enhancement factors using the screening potential energy at the origin \req{eq:weakenhance} are shown as a red dotted line for weak screening and an orange dashed line for strong screening. The orange-shaded region marks the BBN temperature range $T = 50 - 86.2\, $keV.}
    \label{fig:enhanceCarbon}
\end{figure}

As an informative exercise, we show the enhancement factor for $^{12}$C-$^{12}$C fusion in \reff{fig:enhanceCarbon}. This choice of reaction is to connect to stellar fusion in white dwarfs. We can see that at higher $Z$, the strong screening is becoming more important due to its nonlinear dependence on $Z$. Here, the enhancement at large temperatures is compared to weak screening is $\sim 5\times 10^{-4}$. Compared to $^4$He-$^4$He, This is a $\sim 500$ times the stronger effect for an increase in $Z$ of only 4. Although this effect is too small to account for issues in the abundance of light elements, such as the Lithium problem~\citep{Pitrou:2018cgg}, strong screening is very interesting because it can affect reactions with higher Z while leaving low Z reactions unchanged.

The enhancement predicted by strong screening at the origin, shown as the orange dashed line in \reff{fig:enhanceAlpha}, deviates more since the step in the screening potential is more pronounced at higher $Z$. In \reff{fig:enhanceCarbon}, the strong screening enhancement at low temperatures is offset from the weak field limit because the increase in $\phi/T$ compensates for the decrease in the Debye mass.

\section{Conclusion and discussion}\label{sec:Conc_Disc}













\subsection{Summary}
In this study, we implemented self-consistent strong electron-positron plasma screening to determine the short-range screening potential relevant to thermonuclear reactions during the Big Bang nucleosynthesis (BBN) epoch. This research direction was originally proposed in our previous work on BBN screening~\citep{Grayson:2023flr}. Our approach included incorporating finite-sized nuclei into the Poisson-Boltzmann equation and introducing a generalized screening mass to facilitate the comparison of various screening models.

Our results in \reff{fig:enhanceAlpha} and \reff{fig:enhanceCarbon} indicate that strong screening, while generally small due to the high temperatures prevalent during BBN, can enhance the fusion rates of high-$Z$ elements while leaving lower-$Z$ elements relatively unaffected.

One of our key findings is the necessity of using relativistic Fermi-Dirac statistics to describe the polarization cloud accurately at short distances, particularly when the electromagnetic work performed by ions on electrons is comparable to their rest mass energy creating a large polarization density around the nucleus. This is true even when the plasma adheres to Boltzmann statistics at larger distances.

Additionally, we show in \reff{fig:alpha_temp} that the spatial dependence of the strong screening potential near the nuclear surface is very pronounced, having a Fermi-function ``step shape.'' Relying solely on the screening energy near the origin can lead to overestimating reaction rate enhancement due to strong screening.

In this work we accounted for strong screening, the finite size of nuclei, and finite tunneling distances to calculate reaction rate enhancement in the BBN epoch for two simple cases of $^4$He-$^4$He and $^{12}$C-$^{12}$C collisions. These improvements provide a more precise framework for understanding plasma screening effects during BBN and have broader applications in studying weakly coupled plasmas in various cosmic and laboratory settings.

\subsection{Outlook}

Usually, a formulation of strong field dynamics in the semi-classical regime using the Vlasov-Boltzmann equation is intractable due to the nonlinear exponential behavior of the induced charge density in the plasma. We think that the novel analytic form of the Fermi-Dirac screening mass in the ultrarelativistic limit \req{eq:ultrarel} could lead to a solvable strong field theory of plasma dynamics in the high-temperature regime. Our strong screening approach could be relevant to the dynamics of heavy quarks in quark-gluon plasma, where light quarks are ultrarelativistic and strongly screen slow-moving heavier quarks~\citep{mrowczynski2018,CARRINGTON2020}.

Strong screening has recently gained interest in the context of laser plasma fusion, as demonstrated by studies predicting in the context of the Boltzmann limit large enhancements of fusion reactions as compared to weak screening at low temperatures and high densities~\citep{Elsing:2022xyz}. 

Additionally, we are interested in exploring nuclear fusion catalysis by heavy ions. Strong screening predicts a significant screening cloud in proximity to large $Z$ nuclei. Light elements could collide in an environment near the heavy nuclei where their repulsion is mostly neutralized, leading to a consequential enhancement of fusion reactions.

In future studies, we are interested in exploring how the pion exchange impacts the nuclear attraction to refine our understanding of the nuclear charge distribution and the effective tunneling distance. The effective nuclear radius $R$ used in tunneling is an important factor affecting strong screening. Considering a more accurate model of nuclear size would alter the tunneling distance through potential wall. 

Historically, experimental determinations of astrophysical S-factors have found anomalous screening at low temperatures~\citep{Shoppa1993}. This anomalous screening was measured again recently~\citep{Zhang2020ApJ}. We speculate that this anomalous screening at low collision energy could be explained by strong screening effects obtained in this work.

The highly nonlinear response of the strong screening effect to $Z$ suggests that strong screening could help address the observed lithium over-abundance problem by enhancing the fusion of higher-$Z$ elements~\citep{Pitrou:2018cgg}. However, based on the screening enhancements obtained in this study, an explanation of the observed Lithium abundance would require an additional screening mechanism, such as a combined strong-dynamic screening theory.

\section{Acknowledgements}
This research did not receive specific grants from funding agencies in the public, commercial, or not-for-profit sectors. However, we thank Tam\'as Bir\'o for his hospitality at the PP2024 Particles \& Plasmas Symposium, where this work was presented.

\appendix

\section{Solving the Poisson-Boltzmann equation in the weak-field limit}\label{sec:freechg}
In the weak field approximation $Ze\phi/T \ll 1$, the hyperbolic sine term in \req{eq:MassBoltz} can be approximated to first order
\begin{equation}\label{eq:PossBoltzweak}
    -\frac{d^2}{dr^2}\Phi(r)-\frac{2}{r}\frac{d}{dr}\Phi(r)+ \frac{1}{\lambda_D^2} \Phi(r)= P_{\mathrm{ext}}(r)\,,
\end{equation}
for the external charge distribution given in \req{eq:Pext} this has an analytic solution that can be found via Fourier-transform-methods using the boundary condition of vanishing potential for large distances~\citep{Grayson:2023flr}
\begin{equation}\label{eq:Stat_Gauss}
 \phi_{\text{weak}}(r) =\frac{Z e}{4 \pi \varepsilon_0}\frac{e^{ R^2/(4\lambda_D)}}{r}\left(\frac{ \text{Erf}\left(\frac{r}{R} - \frac{R}{2 \lambda_D}\right)}{2 }e^{- r/\lambda_D}+\frac{ \text{Erf}\left(\frac{r}{R} + \frac{ R}{2\lambda_D}\right)}{2 }e^{ r/\lambda_D} - \sinh( r/\lambda_D)\right)\,.
\end{equation}
This solution is valid for large distances where the potential is small enough compared to temperature that the exponential term in \req{eq:denEnhan} is close to one. The vacuum result for $m_D\rightarrow 0$ or $\lambda_D \rightarrow \infty$
\begin{equation}\label{eq:Stat_Vac}
 \phi_{\text{vac}}(r) =\frac{Z e}{4 \pi \varepsilon_0}\frac{\text{Erf}\left(\frac{r}{R}\right)}{r}\,,
\end{equation}
 is often used to model the Coulomb field of $^4$He~\citep{Kumar:2022cdj}.

\section{Checking the numerical solution}\label{sec:check}
One can check for consistency of the solution by integrating the solution to find the total charge and comparing it to the effective charge at large distances represented by the large distance behavior of the solution. To find the total charge predicted by the numerical solution, one rearranges the Poisson equation 
\begin{equation}\label{eq:EffChrg}
    -\nabla^2\Phi(r)+ \frac{m_D^2}{(\hbar c)^2} \Phi(r)=  \left[P_{\mathrm{ext}}(r)+\left(\frac{m_D^2}{(\hbar c)^2}- \frac{m_s^2(r)}{(\hbar c)^2}\right) \Phi(r)\right]\,.
\end{equation}
We then identify the right-hand side as the source for the left-hand side 
\begin{equation} \label{eq:Qeff1}
\begin{split}
Q_{\mathrm{eff}} =  \int dV \Big[P_{\mathrm{ext}}(r)+  \left(\frac{m_D^2}{(\hbar c)^2}- \frac{m_s^2(r)}{(\hbar c)^2}\right) \Phi(r)\Big]\,.
\end{split}
\end{equation}
The large distance solution to the left-hand side is of the Debye-H\"uckel form
\begin{equation}\label{eq:Qeff2}
\Phi_{r \gg \hbar c/m_s} (r)= \frac{Q_{\mathrm{eff}} \,e^{-m_D r}}{r}\,.
\end{equation}
Then the consistency of the numerical solution can be checked by comparing the value of the effective charge found from \req{eq:Qeff1} and \req{eq:Qeff2}.

\section{Enhancement due to finite size}\label{sec:finte}
The textbook formula for the penetration probability of tunneling to some finite distance $R$ at the surface of the nuclei is~\citep{Griffiths:2018}
\begin{equation}
    \mathcal{P}(R) = \exp{\left[ \frac{2 \sqrt{2 \mu_r E_G}}{\hbar c}\left(\frac{\pi}{2}r_c  - 2\sqrt{R r_c}\right)\right]}\,.
\end{equation}
Comparison can also be made with point nuclei charge, where the finite size of the nuclei is neglected, and the tunneling distance goes from the classical turning point to the origin
\begin{equation}\label{eq:SfactorSomer}
    \sigma_{\text{Som}}(E) = \frac{S(E)}{E}\exp{\left[ -2 \pi \eta(E) \right]}\,,
\end{equation}
where $\eta(E)$ in the exponential is often referred to as the Sommerfeld parameter
\begin{equation}
    \eta(E) =Z_1 Z_2 \alpha \sqrt{\frac{\mu_r}{2 E}}\,.
\end{equation}
For the integral \req{eq:integral} using \req{eq:SfactorSomer} one finds the usual
\begin{equation}
I_{\text{Som}} = 4 \sqrt{\frac{2 E_G}{3\mu}} \frac{S(E_G)}{T}e^{-3 E_G /T}\,.
\end{equation}
The ratio of the finite size expression to the Sommerfeld parameter is
\begin{equation}
    \frac{\mathcal{P}(R)}{e^{- 2\pi \eta}} \approx
    \exp \left( 4 \sqrt{2 Z_1 Z_2 \alpha \frac{ R \mu_r}{\hbar c}}\right)\, ,
\end{equation}
which is around 32.8 for alpha-alpha collisions during BBN, a substantial enhancement compared to the effect produced by screening.

\bibliography{main}

\begin{thebibliography}{}
\expandafter\ifx\csname natexlab\endcsname\relax\def\natexlab#1{#1}\fi
\providecommand{\url}[1]{\href{#1}{#1}}
\providecommand{\dodoi}[1]{doi:~\href{http://doi.org/#1}{\nolinkurl{#1}}}
\providecommand{\doeprint}[1]{\href{http://ascl.net/#1}{\nolinkurl{http://ascl.net/#1}}}
\providecommand{\doarXiv}[1]{\href{https://arxiv.org/abs/#1}{\nolinkurl{https://arxiv.org/abs/#1}}}

\bibitem[{Bahcall {et~al.}(2002)Bahcall, Brown, Gruzinov, \& Sawyer}]{Bahcall:2002xyz}
Bahcall, J.~N., Brown, L.~S., Gruzinov, A., \& Sawyer, R.~F. 2002, Astron. Astrophys., 388, 660, \dodoi{10.1051/0004-6361:20020462}

\bibitem[{Bi {et~al.}(2000)Bi, Mauro, \& Christensen-Dalsgaard}]{Bi:2000}
Bi, S.~L., Mauro, M. P.~D., \& Christensen-Dalsgaard, J. 2000, Astron. Astrophys., 364, 157

\bibitem[{Birrell {et~al.}(2024)Birrell, Formanek, Steinmetz, Yang, \& Rafelski}]{Birrell:2024bdb}
Birrell, J., Formanek, M., Steinmetz, A., Yang, C.~T., \& Rafelski, J. 2024, International Journal of Theoretical Physics.
\newblock \doarXiv{2405.05287}

\bibitem[{Br\"uggen \& Gough(1997)}]{Bruggen:1997}
Br\"uggen, M., \& Gough, D.~O. 1997, Astrophys. J., 488, 867, \dodoi{10.1086/304718}

\bibitem[{Br\"uggen \& Gough(2000)}]{Bruggen:2000}
---. 2000, Journal of Mathematical Physics, 41, 260

\bibitem[{Carraro {et~al.}(1988)Carraro, Schafer, \& Koonin}]{Carraro:1988apj}
Carraro, C., Schafer, A., \& Koonin, S.~E. 1988, Astrophys. J., 331, 565, \dodoi{10.1086/166582}

\bibitem[{Carrington {et~al.}(2020)Carrington, Czajka, \& Mrówczyński}]{CARRINGTON2020}
Carrington, M.~E., Czajka, A., \& Mrówczyński, S. 2020, Nuclear Physics A, 1001, 121914, \dodoi{https://doi.org/10.1016/j.nuclphysa.2020.121914}

\bibitem[{Cowan \& Kirkwood(1958)}]{Cowan:1958xyz}
Cowan, R.~D., \& Kirkwood, J.~G. 1958, Phys. Rev., 111, 1460, \dodoi{10.1103/PhysRev.111.1460}

\bibitem[{Debye \& Hückel(1923)}]{Debye:1923}
Debye, P., \& Hückel, E. 1923, Physikalische Zeitschrift, 24, 185–206

\bibitem[{Dewitt {et~al.}(1973)Dewitt, Graboske, \& Cooper}]{Dewitt1973}
Dewitt, H.~E., Graboske, H.~C., \& Cooper, M.~S. 1973, Astrophysical Journal, 181, 439

\bibitem[{Dzitko {et~al.}(1995)Dzitko, Turck-Chieze, Delbourgo-Salvador, \& Lagrange}]{Dzitko:1995xyz}
Dzitko, H., Turck-Chieze, S., Delbourgo-Salvador, P., \& Lagrange, C. 1995, Astrophys. J., 447, 428, \dodoi{10.1086/175887}

\bibitem[{Elsing {et~al.}(2022)Elsing, P\'alffy, \& Wu}]{Elsing:2022xyz}
Elsing, D., P\'alffy, A., \& Wu, Y. 2022, Phys. Rev. Res., 4, L022004, \dodoi{10.1103/PhysRevResearch.4.L022004}

\bibitem[{Elze {et~al.}(1980)Elze, Greiner, \& Rafelski}]{Elze:1980er}
Elze, H.~T., Greiner, W., \& Rafelski, J. 1980, J. Phys. G, 6, L149, \dodoi{10.1088/0305-4616/6/9/003}

\bibitem[{Famiano {et~al.}(2016)Famiano, Balantekin, \& Kajino}]{Famiano:2016hhs}
Famiano, M.~A., Balantekin, A.~B., \& Kajino, T. 2016, Phys. Rev. C, 93, 045804, \dodoi{10.1103/PhysRevC.93.045804}

\bibitem[{Fogolari {et~al.}(2002)Fogolari, Brigo, \& Molinari}]{Fogolari:2002xyz}
Fogolari, F., Brigo, A., \& Molinari, H. 2002, J. Mol. Recognit., 15, 377

\bibitem[{Formanek {et~al.}(2021)Formanek, Grayson, Rafelski, \& M\"uller}]{Formanek:2021blc}
Formanek, M., Grayson, C., Rafelski, J., \& M\"uller, B. 2021, Annals Phys., 434, 168605, \dodoi{10.1016/j.aop.2021.168605}

\bibitem[{Frolov(2023)}]{Frolov:2023zkf}
Frolov, A.~M. 2023, Phys. Plasmas, 30, 102701, \dodoi{10.1063/5.0156153}

\bibitem[{Graboske {et~al.}(1973)Graboske, Dewitt, Grossman, \& Cooper}]{Graboske:1973xyz}
Graboske, H.~C., Dewitt, H.~E., Grossman, A.~S., \& Cooper, M.~S. 1973, Astrophys. J., 181, 457, \dodoi{10.1086/152062}

\bibitem[{Grayson {et~al.}(2023)Grayson, Yang, Formanek, \& Rafelski}]{Grayson:2023flr}
Grayson, C., Yang, C.~T., Formanek, M., \& Rafelski, J. 2023, Annals Phys., 458, 169453, \dodoi{10.1016/j.aop.2023.169453}

\bibitem[{Griffiths \& Schroeter(2018)}]{Griffiths:2018}
Griffiths, D.~J., \& Schroeter, D.~F. 2018, Introduction to Quantum Mechanics (Cambridge University Press)

\bibitem[{Groot {et~al.}(1980)Groot, Leeuwen, \& Weert}]{DeGroot:1980dk}
Groot, S. R.~D., Leeuwen, W. A.~V., \& Weert, C. G.~V. 1980, Relativistic Kinetic Theory. Principles and Applications (North-Holland Publishing Company)

\bibitem[{Gruzinov(1998)}]{Gruzinov:1997as}
Gruzinov, A.~V. 1998, Astrophys. J., 496, 503, \dodoi{10.1086/305349}

\bibitem[{Gruzinov \& Bahcall(1998)}]{Gruzinov:1998}
Gruzinov, A.~V., \& Bahcall, J.~N. 1998, Astrophys. J., 504, 996, \dodoi{10.1086/306116}

\bibitem[{Hakim(1967)}]{Hakim:1967prd}
Hakim, R. 1967, Phys. Rev., 162, 128, \dodoi{10.1103/PhysRev.162.128}

\bibitem[{Hwang {et~al.}(2021)Hwang, Jang, Park, Kusakabe, Kajino, Balantekin, Maruyama, Ryu, \& Cheoun}]{Hwang:2021kno}
Hwang, E., Jang, D., Park, K., {et~al.} 2021, JCAP, 11, 017, \dodoi{10.1088/1475-7516/2021/11/017}

\bibitem[{Ichimaru(1982)}]{Ichimaru:1982}
Ichimaru, S. 1982, Rev. Mod. Phys., 54, 1017, \dodoi{10.1103/RevModPhys.54.1017}

\bibitem[{Itoh {et~al.}(1977)Itoh, Totsuji, \& Ichimaru}]{Itoh1977}
Itoh, N., Totsuji, H., \& Ichimaru, S. 1977, Astrophysical Journal, 218, 477, \dodoi{10.1086/155701}

\bibitem[{Itoh {et~al.}(1979)Itoh, Totsuji, Ichimaru, \& DeWitt}]{Itoh1979}
Itoh, N., Totsuji, H., Ichimaru, S., \& DeWitt, H.~E. 1979, Astrophysical Journal, 234, 1079, \dodoi{10.1086/157590}

\bibitem[{Kedia {et~al.}(2021)Kedia, Sasankan, Mathews, \& Kusakabe}]{Kedia:2020xdc}
Kedia, A., Sasankan, N., Mathews, G.~J., \& Kusakabe, M. 2021, Phys. Rev. E, 103, 032101, \dodoi{10.1103/PhysRevE.103.032101}

\bibitem[{Kodama(2002)}]{Kodama:2002}
Kodama, T. 2002, in AIP Conference Proceedings, Vol. 631 (AIP), 3--26

\bibitem[{Krauth {et~al.}(2021)Krauth, Schuhmann, Ahmed, Amaro, Amaro, Biraben, Chen, Covita, Dax, Diepold, {et~al.}}]{Krauth:2021foz}
Krauth, J.~J., Schuhmann, K., Ahmed, M.~A., {et~al.} 2021, Nature, 589, 527, \dodoi{10.1038/s41586-021-03183-1}

\bibitem[{Kravchuk \& Yakovlev(2014)}]{Kravchuk:2014xsa}
Kravchuk, P.~A., \& Yakovlev, D.~G. 2014, Phys. Rev. C, 89, 015802, \dodoi{10.1103/PhysRevC.89.015802}

\bibitem[{Kumar {et~al.}(2022)Kumar, Awasthi, Khachi, \& Sastri}]{Kumar:2022cdj}
Kumar, L., Awasthi, S., Khachi, A., \& Sastri, O. S. K.~S. 2022.
\newblock \doarXiv{2209.00951}

\bibitem[{Liolios(2004)}]{Liolios:2004}
Liolios, T.~E. 2004, Eur. Phys. J. A, 18, s1, \dodoi{10.1140/epjad/s2003-01-001-7}

\bibitem[{Luo {et~al.}(2020)Luo, Famiano, Kajino, Kusakabe, \& Balantekin}]{Luo:2020}
Luo, Y., Famiano, M.~A., Kajino, T., Kusakabe, M., \& Balantekin, A.~B. 2020, Phys. Rev. D, 101, 083010, \dodoi{10.1103/PhysRevD.101.083010}

\bibitem[{Mr{\'o}wczy{\'n}ski(2018)}]{mrowczynski2018}
Mr{\'o}wczy{\'n}ski, S. 2018, The European Physical Journal A, 54, 1

\bibitem[{Opher \& Opher(2000)}]{Opher:1999jh}
Opher, M., \& Opher, R. 2000, Astrophys. J., 535, 473, \dodoi{10.1086/308808}

\bibitem[{Pitrou {et~al.}(2018)Pitrou, Coc, Uzan, \& Vangioni}]{Pitrou:2018cgg}
Pitrou, C., Coc, A., Uzan, J.~P., \& Vangioni, E. 2018, Phys. Rept., 754, 1, \dodoi{10.1016/j.physrep.2018.04.005}

\bibitem[{Rafelski {et~al.}(2023)Rafelski, Birrell, Steinmetz, \& Yang}]{Rafelski:2023emw}
Rafelski, J., Birrell, J., Steinmetz, A., \& Yang, C.~T. 2023, Universe, 9, 309, \dodoi{10.3390/Universe9070309}

\bibitem[{Rose(1998)}]{Rose1998}
Rose, W.~K. 1998, Advanced Stellar Astrophysics (Cambridge University Press)

\bibitem[{Salpeter(1954)}]{Salpeter:1954nc}
Salpeter, E.~E. 1954, Austral. J. Phys., 7, 373, \dodoi{10.1071/PH540373}

\bibitem[{Salpeter \& van Horn(1969)}]{Salpeter:1969apj}
Salpeter, E.~E., \& van Horn, H.~M. 1969, Astrophys. J., 155, 183, \dodoi{10.1086/149858}

\bibitem[{Sasankan {et~al.}(2020)Sasankan, Kedia, Kusakabe, \& Mathews}]{Sasankan:2019oee}
Sasankan, N., Kedia, A., Kusakabe, M., \& Mathews, G.~J. 2020, Phys. Rev. D, 101, 123532, \dodoi{10.1103/PhysRevD.101.123532}

\bibitem[{Shaviv \& Shaviv(1996)}]{Shaviv:1996}
Shaviv, N.~J., \& Shaviv, G. 1996, Astrophys. J., 468, 433, \dodoi{10.1086/177702}

\bibitem[{Shoppa {et~al.}(1993)Shoppa, Koonin, Langanke, \& Seki}]{Shoppa1993}
Shoppa, T.~D., Koonin, S.~E., Langanke, K., \& Seki, R. 1993, Phys. Rev. C, 48, 837, \dodoi{10.1103/PhysRevC.48.837}

\bibitem[{Smith {et~al.}(2020)Smith, Bishop, Hirst, {et~al.}}]{smith2020hoyle}
Smith, R., Bishop, J., Hirst, J., {et~al.} 2020, Few-Body Systems, 61, 14, \dodoi{10.1007/s00601-020-1545-5}

\bibitem[{Wang {et~al.}(2011)Wang, Bertulani, \& Balantekin}]{Wang:2010px}
Wang, B., Bertulani, C.~A., \& Balantekin, A.~B. 2011, Phys. Rev. C, 83, 018801, \dodoi{10.1103/PhysRevC.83.018801}

\bibitem[{Yao {et~al.}(2017)Yao, Mehen, \& M\"uller}]{Yao:2016cjs}
Yao, X., Mehen, T., \& M\"uller, B. 2017, Phys. Rev. D, 95, 116002, \dodoi{10.1103/PhysRevD.95.116002}

\bibitem[{{Zhang} {et~al.}(2020){Zhang}, {Huang}, {Hu}, {Chen}, {Hou}, {Wang}, \& {Fang}}]{Zhang2020ApJ}
{Zhang}, Q., {Huang}, Z., {Hu}, J., {et~al.} 2020, \apj, 893, 126, \dodoi{10.3847/1538-4357/ab8222}

\end{thebibliography}

\end{document}